\documentstyle[11pt,newpasp,twoside,psfig]{article}
\def\ltsima{$\; \buildrel < \over \sim \;$}
\def\simlt{\lower.5ex\hbox{\ltsima}}
\def\gtsima{$\; \buildrel > \over \sim \;$}
\def\simgt{\lower.5ex\hbox{\gtsima}}
\begin{document}

\title{LINERs and their Physical Mechanisms}
\author{Alexei V. Filippenko}
\affil{Department of Astronomy, University of California, Berkeley, 
CA 94720-3411, USA~~~
email: alex@astro.berkeley.edu}

\setcounter{page}{1}
% this page number will be filled later by the editors....
\index{Filippenko, A. V.}

\begin{abstract}

I review the basic properties of low-ionization nuclear emission-line regions
(LINERs), as well as the main ionization mechanisms thus far proposed for
them. There is substantial heterogeneity among LINERs, especially those with
extended kpc-scale emission. An extensive survey by Ho, Filippenko, \& Sargent
quantifies with unprecedented accuracy the optical emission-line intensity
ratios in the nuclei of a complete sample of 486 nearby, bright galaxies.  Over
40\% of all early-type galaxies in the sample can be classified as LINERs or
LINER/H~II transition objects. A nonstellar continuum is detected in many
LINERs, suggesting that they are genuine AGNs powered by accretion onto a
compact object. Broad H$\alpha$ emission similar to that of Seyfert 1 nuclei is
seen in many LINERs. In some LINERs, the broad H$\alpha$ emission has two
well-separated peaks, probably indicative of an accretion disk.

\end{abstract}

\section{Introduction}

   Over 20 years ago, Heckman (1980) described a class of galactic nuclei whose
optical spectra are quite distinct from those of both H~II regions and
classical active galactic nuclei (AGNs). These objects, dubbed ``Low-Ionization
Nuclear Emission-line Regions'' (LINERs), are characterized by narrow emission
lines of relatively low ionization.  Specifically, membership in this class was
defined by only two line-intensity ratios: [O~II] $\lambda$3727/[O~III]
$\lambda$5007 $\ge 1$, and [O~I] $\lambda$6300/[O~III] $\lambda$5007 $\ge
1/3$. Since the exact divisions are somewhat arbitrary, objects formally
satisfying only one of the criteria, but nearly consistent with the other, are
also commonly called LINERs.

   Several authors (e.g., Veilleux \& Osterbrock 1987) noticed that unambiguous
LINERs, such as NGC 1052, have [N~II] $\lambda$6583/H$\alpha$ $\simgt 0.6$ and
[O~III] $\lambda$5007/H$\beta$ $\simlt 3$, unlike most H~II regions.  Kewley et
al. (2001) adopted a related scheme using theoretical models.  Also, some
samples of LINERs in the literature have relied on only the [O~II]/[O~III]
ratio; [O~I] was not visible due to low signal-to-noise ratios or insufficient
spectral coverage.

   Heckman (1980) estimated that $\simgt 1/3$ of all spiral galaxies are
LINERs, and other surveys of nearby galaxies also showed that LINERs are very
common, especially in early-type galaxies (Ho, Filippenko, \& Sargent 1997a,
and references therein). Given their great preponderance, it is important to
determine whether LINERs are miniature (very low luminosity) QSOs; if so, we
should certainly include them in the luminosity function of AGNs.

    Although many low-luminosity ``AGNs" (in quotes, because we still don't
know their exact physical nature) are LINERs, there are some exceptions,
such as the {\it Seyfert 1} nucleus of NGC 4395 (Filippenko, Ho, \& Sargent
1993).  Thus, ``low luminosity" does not necessarily imply ``low ionization."
Also, some luminous ``AGNs" are LINERs, especially bright infrared galaxies.

  Note that extended (few kpc) ``LINER" emission is sometimes seen in
galaxies having starbursts, galactic winds, or cooling flows, or in 
interacting/merging galaxies. These objects are {\it not} the focus of
the current review, although the relevant physical mechanisms will be
considered to some extent. Rather, the emphasis here is on LINERs 
within a radius of 200 pc from the nucleus.

%\begin{figure}[1]
\begin{center}
\psfig{figure=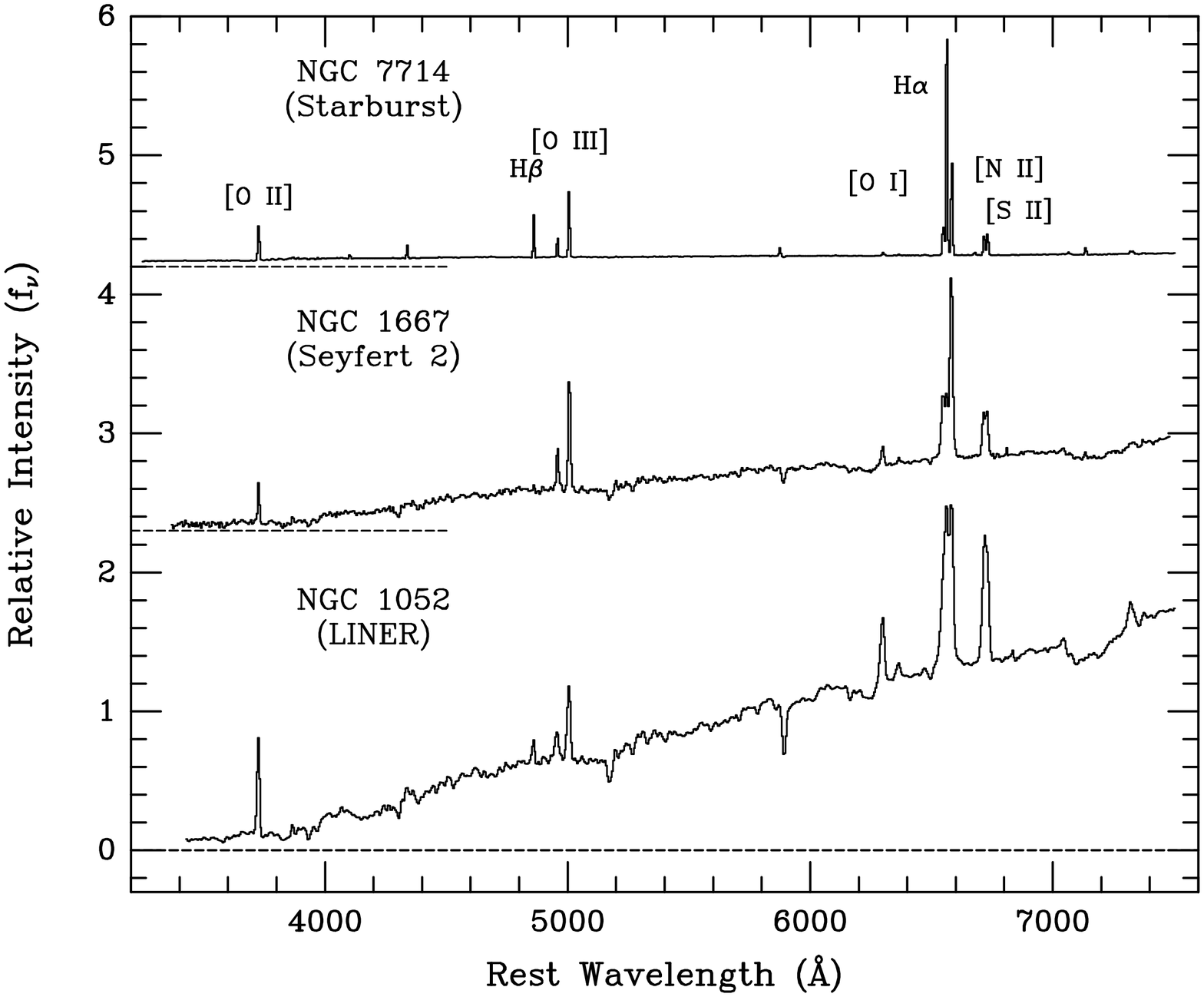,width=6cm,angle=-90}
\end{center}
\vskip 6.5 truecm
\noindent {\bf Fig. 1.}
{Spectra of NGC 7714 (a normal starburst galaxy), NGC 1667 (a
Seyfert 2 galaxy), and NGC 1052 (a LINER). The relative strengths of
the oxygen emission lines are used by Heckman (1980) to define LINERs.}
%\label{fig1}
%\end{figure}
\bigskip

\section{The Palomar Observatory Galaxy Survey}

   An extensive survey of 486 bright, nearby galaxies that my collaborators and
I conducted now makes it possible to reassess the demographics of LINERs and to
determine in detail their optical characteristics. The sample consists of all
galaxies with $B_T \leq 12.5$ mag and $\delta > 0^\circ$ (Filippenko \& Sargent
1985). CCD spectra were obtained with the 5-m Hale reflector at Palomar
Observatory over the wavelength ranges 4230--5110~\AA\ and 6210--6860~\AA\ with
roughly 4~\AA\ and 2.5~\AA\ resolution, respectively. An atlas of spectra is
presented by Ho, Filippenko, \& Sargent (1995). Excellent starlight subtraction
was achieved through a $\chi^2$ minimization algorithm (Rix \& White 1982)
involving the large number ($\sim 60$) of absorption-line ``template" galaxies
in the survey.  We were consequently able to measure the relative intensities
of most lines to high accuracy (Ho, Filippenko, \& Sargent 1997a), permitting
detailed comparisons with calculations of excitation mechanisms.

   We confirm that LINERs are very common (Ho, Filippenko, \& Sargent 1997b),
especially among early-type galaxies. An important distinction, however, is
that our results are quantitatively more reliable than those of previous
surveys, both in a statistical sense and for individual objects. This is due to
our combination of small entrance apertures, moderate resolution, high
signal-to-noise ratios, and careful starlight subtraction. If we define the
LINER category to include both ``pure" LINERs and ``transition objects" (those
with low [O~I] $\lambda$6300/H$\alpha$ ratios; they may be LINER/H~II
composites), then at least 32\% of all bright galaxies are LINERs. LINERs are
found among $\sim 40$\% of E, S0, and Sb galaxies, and among 60\% of Sa
galaxies. For comparison, none of the E--E/S0 galaxies in our sample contain
H~II nuclei; instead, H~II nuclei reside in over 75\% of late-type galaxies (Sc
and later). If we temporarily assume that all LINERs and transition objects are
indeed AGNs, then over 50\% of E, S0, and Sb galaxies contain active nuclei;
the fraction is closer to 75\% for Sa galaxies. Of the complete sample of 486
galaxies in the Palomar survey, 43\% are AGNs: 11\% Seyfert, 19\% LINERs, and
13\% transition objects.

\section{LINERS: True AGNs or Impostors?}

\subsection{Photoionization by a Nonstellar Continuum}

   Based on the spectroscopic resemblance between LINERs and supernova
remnants, as well as on high electron temperatures ($T_e$) derived from the
[O~III] $\lambda$4363/[O~III] $\lambda$5007 intensity ratio, several early
studies concluded that the emission lines in LINERs are produced by
shock-heated gas (Koski \& Osterbrock 1976; Fosbury et al. 1978; Heckman 1980),
which may have nothing to do with accretion onto a compact object. In fact,
Baldwin, Phillips, \& Terlevich (1981) used the line-intensity ratios observed
in LINERs to {\it define} the region populated by ``shock-heated galaxies'' in
their two-dimensional classification schemes (see also Veilleux \& Osterbrock
1987).

   However, the apparent continuity of LINERs and Seyfert galaxies in the
Baldwin et al. (1981) diagrams, together with the discovery of significant
X-ray emission in some LINERs, soon led to the development of power-law
photoionization models of LINERs (Ferland \& Netzer 1983; Halpern \& Steiner
1983). In this view, LINERs are genuine AGNs, sustained by whatever mechanism
produces the nonstellar UV continuum in QSOs (presumably accretion of matter
onto a supermassive black hole), but with a low ionization parameter $U$ (ratio
of ionizing photons to nucleons at the face of a gas cloud).

    Two problems with this interpretation were that the derived value of $T_e$
was too high in a few objects, and that the entire {\it set} of emission-line
intensity ratios in a given object could be difficult to match with low-density
models (e.g., the [O~I] $\lambda$6300 line was sometimes observed to be very
strong). The discovery of a wide range of electron densities ($n_e =
10^2$--$10^7$ cm$^{-3}$) in the narrow-line regions of some well-studied
LINERs, based on the presence of a strong correlation between line width and
critical density for de-excitation, removed or substantially ameliorated these
problems (Filippenko \& Halpern 1984; Filippenko 1985; Filippenko \& Sargent
1988; Ho, Filippenko, \& Sargent 1993) --- although this is still occasionally
overlooked in the literature (e.g., Dopita \& Sutherland 1995; Kewley et
al. 2001). For example, the value of $T_e$ deduced from the [O~III] lines
decreases considerably when collisional suppression of the [O~III]
$\lambda\lambda$4959, 5007 lines relative to [O~III] $\lambda$4363 is properly
taken into account.

  Most recently, Barth et al. (2001) analyzed {\it Hubble Space Telescope
(HST)} spectra of the LINER NGC 4579 that directly demonstrate the presence of
a density gradient within the inner arcsecond of the narrow-line region.  The
intensity ratio of the narrow [S~II] $\lambda\lambda$6716, 6731 lines exhibits
a pronounced gradient, indicating a steep rise in density toward the nucleus.
Moreover, the [O~I] $\lambda$6300/[S~II] $\lambda\lambda$6716, 6731 ratio
increases sharply with decreasing distance from the nucleus. There is now no
doubt that the narrow-line region is stratified, and that a range of densities
must be considered. When this is done, the ``high-$T_e$ problem'' vanishes; the
temperature in the O$^{++}$ zone is comfortably below 20,000~K, and
photoionization models work well for compact LINERs.

\subsection{Shocks}

   Shock heating is clearly important in some {\it extended} LINERs --- those
involving superwinds, interactions, mergers, and radio jets entraining
ambient material. For example, NGC 3079 (Filippenko \& Sargent 1992) exhibits
ample evidence of shock-excited gas, as do many IRAS galaxies. The relevance
of shocks to galactic {\it nuclei}, on the other hand, is less obvious.

   Dopita \& Sutherland (1995, 1996) have revitalized interest in shock models
of Seyferts and LINERs; the key new insights are that the shock velocities and
magnetic field strengths may be very large.  High-velocity shocks ($v \approx
200$--500 km s$^{-1}$) produce energetic photons capable of ionizing unshocked
upstream gas; the resulting emission-line intensity ratios are similar to those
observed. Dopita \& Sutherland postulate that in Seyfert nuclei, there is
plenty of gas to be photoionized, while in LINERs there is a dearth of gas (and
hence the high-ionization lines are weak).

   High-velocity shocks may well be operating in the circumnuclear region of
the LINER M87: {\it HST}~ UV/optical spectra of part of the central disk show
strong emission lines having relative intensities in good agreement with
theoretical predictions (Dopita et al. 1997). However, the UV and optical
emission lines in the {\it exact nucleus} of M87 have intensity ratios that are
inconsistent with the predictions of shock models (Sabra et al. 2003; Shields
et al., these proceedings). Furthermore, the emission lines in most LINERs and
Seyferts have FWHM $\approx 100$--200 km s$^{-1}$, too slow for the relevant
shocks. It is also energetically difficult for shocks to produce the amount of
line emission seen in galactic nuclei; the gas must be repeatedly
re-energized. The clear presence of a point-like source of ionizing radiation
in some LINERs and Seyfert 2 nuclei, and in most type 1 Seyferts, does not
appear to be explained in the current shock models.  Finally, as mentioned by
Andrew Wilson at this meeting, the {\it Chandra} X-ray Observatory has not
revealed the presence of very hot gas in some key objects such as the Seyfert
nucleus of NGC 1068. I conclude that shock heating may be dominant in some
nuclear {\it regions}, but probably not in many galactic {\it nuclei}. Rather,
accretion onto a compact, supermassive black hole is more likely to dominate.

\subsection{Cooling Flows}

   Another alternative is that some LINERs may be cooling accretion flows
(e.g., Heckman 1981; Fabian et al. 1986; Heckman et al. 1989), since optical
spectra of LINERs closely resemble those of extended gas in some giant
elliptical galaxies such as M87. Indeed, cursory inspection of Figure 2a in
Shields \& Filippenko (1990) shows that the filaments of NGC 1275 (in the
Perseus cluster) easily meet the defining characteristics of LINERs. Sabra,
Shields, \& Filippenko (2000, and references therein) discuss several possible
mechanisms to explain the energetics and spectra of the NGC 1275 filaments, but
no clear picture has emerged. 

    Long-slit spectra of some early-type galaxies demonstrate that the
emission-line intensity ratios are not very dependent on projected radial
distance from the galactic nucleus (Filippenko 1984). This is very difficult to
understand in the context of photoionization models, since $U \propto
(nr^2)^{-1}$, and the density appears to be independent of radial distance ($r$)
in these objects. It might, however, be consistent with the cooling-flow
hypothesis. On the other hand, {\it most} LINERs are probably not cooling-flow
galaxies; we don't see extended emission-line gas of this type.

\subsection{Stellar Photoionization}

   Hot, evolved, low-mass stars (post-asymptotic giant branch stars) might
provide sufficient ionizing photons to produce the weak, extended [N~II]
(LINER-like) emission lines observed in some galaxies (Binette et al. 1994;
Taniguchi, Shioya, \& Murayama 2000). However, this is probably not applicable
to most {\it compact} LINERs.

   The ``Warmer" hypothesis of Terlevich \& Melnick (1985) has received much
attention in the past, and it may be relevant in a few galactic nuclei
classified as Seyferts and LINERs.  However, {\it HST} spectra of Seyferts and
LINERs don't reveal the presence of Wolf-Rayet stars in almost all cases (e.g.,
Barth et al., these proceedings). A less extreme variant of this model is that
O-star photoionization powers some LINERs, especially the transition objects
(Filippenko \& Terlevich 1992) and galactic nuclei having high electron
densities (Shields 1992). Motivated by the UV spectrum of the LINER NGC 4569,
which clearly reveals a recent starburst, extensive stellar photoionization
models involving realistic star clusters have been explored by Barth \& Shields
(2000).

   Very recently, Ho, Filippenko, \& Sargent (2003) tested the stellar
photoionization hypotheses by examining the stellar continua and absorption
lines in their Palomar survey of nearby galactic nuclei. They find that LINERs
and Seyferts generally have {\it old} stellar populations; there is no clear
evidence for young stars. This applies even to LINER/H~II transition objects,
although enhanced star formation is sometimes visible. Similarly, Barth et
al. (these proceedings) examined {\it HST} STIS spectra of 13 transition
objects and generally found an old stellar population. While some of the
objects are true LINERs on very small angular scales, others retain their
``transition" classification. These data do not firmly support the stellar
photoionization hypothesis in compact galactic nuclei; however, {\it extended}
starbursts and super-starbursts are clearly relevant in many IRAS galaxies and
similar objects (e.g., Kewley et al. 2001).

\section{Tests of the AGN Hypothesis}

  As discussed above, there is now much circumstantial evidence that many
(though not all) compact LINERs are powered by accretion, as are QSOs and
genuine AGNs. The discovery that supermassive black holes are present in the
nuclei of a majority of massive, nearby galaxies (e.g., Gebhardt et al. 2000)
is consistent with this conclusion. Ironically, owing to an observational
selection bias, the galaxies having the best existing dynamical evidence for
supermassive black holes tend to be LINERs and low-luminosity AGNs whose
spectra are {\it not} dominated by emission lines. But there are other,
independent tests of the AGN hypothesis for LINERs, as discussed below.

\subsection{Broad H$\alpha$ Emission}

   An important aspect of our Palomar survey is the search for broad H$\alpha$
emission similar to, but weaker than that found in classical type 1 Seyferts
and QSOs. Indeed, one of our main goals is to quantify the luminosity function
of broad-lined AGNs. Identification of broad H$\alpha$ emission in LINERs
(``LINER 1s") is especially significant, as these objects then become prime
candidates for genuine AGNs. [Before continuing, I note that some authors
(e.g., Krolik 1999) incorrectly state that LINERs having broad permitted
emission lines should be called Seyfert galaxies. Actually, however, the
LINER/Seyfert classification is based on the intensity ratios of the {\it
narrow} components of emission lines. The presence or absence of broader
permitted lines is used to subclassify an object as ``type 1" and ``type 2,"
respectively.]

    Our search technique and conclusions are described in detail by Ho et
al. (1997c). We find that 24\% of ``pure" LINERs contain broad H$\alpha$
emission --- a lower limit to the true fraction of LINER 1s, since the feature
may sometimes be below our level of detectability. Very few (about 4\%) of the
transition objects exhibit broad H$\alpha$ emission.  The Hubble types of the
galaxies are Sbc and earlier, and the fraction is highest among E--E/S0
LINERs. Type 1 AGNs are found in 9.5\% of the Palomar galaxies; they constitute
22\% of all AGNs (16\% of all LINERs including transition, and 37\% of all
Seyferts). In the {\it HST} STIS survey of transition objects by Barth et al.
(these proceedings), 1 new broad-lined object was found among the 13 observed
galaxies.

   It is possible that the ``unified model" for Seyfert galaxies (Antonucci
1993), in which a majority of type 2 Seyferts are obscured Seyfert 1s, also
extends to LINERs. To test this hypothesis, we have used the Keck 10-m
telescope to conduct a spectropolarimetric survey of low-luminosity AGNs
(LINERs and low-luminosity Seyferts). We uncovered the first known examples of
hidden broad-line regions in LINERs (including the prototypical LINER, NGC
1052), demonstrating that at least some of these objects are genuine AGNs
surrounded by obscuring tori, analogous to the torus model for Seyfert nuclei
and high-power radio galaxies (Barth et al. 1999a,b). 

\subsection{The Nonstellar Continuum}

  Another way to test the AGN hypothesis is to search for a point-like
nonstellar continuum in LINERs. Since the optical continuum of LINERs is
generally dominated by old, red stars, this is much more easily done at UV
wavelengths. The surveys by Maoz et al. (1995) and Barth et al. (1998) revealed
that only $\sim 25$\% of LINERs harbor a compact source of UV emission. Dust
probably blocks the nucleus in many cases, as suggested by the fact that the
average galaxy disk inclinations of ``UV-bright" and ``UV-dark" nuclei are
$36^\circ$ and $63^\circ$, respectively. The fraction of intrinsically
UV-bright LINERs may therefore be closer to 50\%. However, to determine the
physical nature of the UV source, spectra are needed.

  The first high-quality {\it HST}~ UV spectrum of a LINER was that of M81 (Ho,
Filippenko, \& Sargent 1996). It exhibits broad permitted emission lines,
further reinforcing the similarity to Seyfert 1 nuclei. A featureless
continuum, most likely of nonstellar origin, is detected unambiguously for the
first time in M81. A UV spectrum of the LINER 1 NGC 4579 is similar to that of
M81 (Barth et al. 1996), and several other LINER 1s also broadly resemble
Seyfert 1s in the UV. On the other hand, some LINER 2s exhibit evidence for
young starbursts (e.g., NGC 4569; Maoz et al. 1998), while the data for others are
still ambiguous.

  Perhaps the clearest signature thus far for AGNs in many LINERs is provided
by X-ray images (e.g., Roberts \& Warwick 2000; Halderson et al. 2001). The
most recent evidence is the most compelling: an unresolved hard X-ray core is
found in 75\% of the 24 galaxies in the high-resolution {\it Chandra} HRI
survey of Ho et al. (2001), including all LINER 1s, most LINER 2s, and a few
transition objects. This strongly suggests that LINER 1s and most LINER 2s are
genuine AGNs. In addition, ASCA measurements show that the X-ray luminosity is
proportional to the H$\alpha$ luminosity (Terashima, Ho, \& Ptak 2000), as
previously found for more luminous AGNs.

  Additional strong evidence for nonstellar processes in LINERs comes from
radio imaging surveys (e.g., Nagar et al. 2000; Filho, Barthel, \& Ho 2000,
2002): 64\% of LINER 1s and 36\% of LINER 2s have compact radio cores. Ho et
al. (2003, in prep.) show that 80\% of LINERs have a flat-spectrum radio core,
with LINER 1s outnumbering LINER 2s in this regard. Only 25\% of transition
objects have such a core, while 0\% of H~II galaxies exhibit one. Complementing
these results, the VLBI measurements of some LINER 1s give brightness
temperatures exceeding $10^8$~K (Falcke et al. 2000), a clear sign of
nonstellar activity.

\section{Double-Peaked Broad H$\alpha$ Emission}

  Quite a few LINERs exhibit very broad, double-peaked H$\alpha$ emission.  The
prototypical case is that of Arp 102B (Halpern et al. 1996, and references
therein; Ho et al. 2000; Shields et al. 2000). It has been shown by Eracleous
\& Halpern (1994) that AGNs exhibiting such profiles also distinguish
themselves in several other respects. Specifically, nearly all of them are
radio loud, and they tend to (a) be dominated by a stellar continuum at optical
wavelengths, (b) have Balmer lines which are about twice as broad as those of
typical radio-loud AGNs, and (c) have unusually strong, low-ionization
forbidden lines whose intensity ratios are reminiscent of those of
LINERs. Additional possible properties may include a flat far-infrared spectrum
and a weak UV continuum, with little evidence for the ``big blue bump."

   Halpern and collaborators have considered in some detail the possibility
that the line profiles are produced by an accretion disk (Chen, Halpern, \&
Filippenko 1989; Eracleous \& Halpern 1994). The {\it HST} spectra of Arp 102B
presented by Halpern et al. (1996) strongly support the application of the
accretion disk model of Dumont \& Collin-Souffrin (1990a,b,c), in which the
disk produces only low-ionization lines. Moreover, the work of Ho, Filippenko,
\& Sargent (1996), Ho (1999, and these proceedings), and others suggests that
low-luminosity AGNs accreting at low rates relative to the Eddington rate have
geometrically thick, ion-supported inner tori, providing a natural explanation
for many of the observed properties of double-peaked AGNs.

\section{Conclusions}

   I summarize this review by providing a ``score card" for galactic nuclei.
The first trait is the presence of a supermassive black hole. By examining
lists of specific objects in papers having dynamical mass estimates, I conclude
that such an object exists in most or all LINER 1s, and usually in LINER 2s as
well. It is sometimes (often?) present in transition objects, and perhaps
rarely in H~II nuclei, although definitive conclusions cannot yet be made
because there are too few good measurements of late-type galaxies (the most
common H~II and transition objects). The second trait, broad H$\alpha$
emission, is present in all LINER 1s (by definition!), some LINER 2s (in
polarized flux, but occasionally barely visible in total flux as well), rarely
in transition objects, and never in H~II nuclei.  Third, a point source
emitting a nonstellar continuum (UV, X-ray, radio) is almost always present in
LINER 1s, often (usually?) in LINER 2s, rarely in transition objects, and never
in H~II nuclei. Fourth, the stellar population is nearly always old in LINER 1s
and LINER 2s, usually old (but sometimes young) in transition objects, and
always young in H~II nuclei. Fifth, the narrow-line gas is photoionized by a
nonstellar continuum in all LINER 1s and most LINER 2s, while gas in transition
nuclei can be photoionized by nonstellar or (sometimes) stellar continua, and
H~II nuclei are photoionized by hot stars.

  From these observed characteristics, I conclude that LINER 1s are always, or
almost always, genuine AGNs. Similarly, LINER 2s are often (perhaps usually)
genuine AGNs. Transition objects are a mixture of AGNs and other processes
(generally starbursts?); in some cases they may be hidden AGNs that will be
revealed by more detailed searches.  H~II nuclei are powered by starbursts,
regardless of whether a central black hole is present. These conclusions apply
to compact nuclei, not to extended regions whose emission lines are produced by
shocks, superwinds, and cooling flows.

\begin{acknowledgments}

Much of my own work presented here was obtained in collaboration with Aaron
Barth, Mike Eracleous, Jules Halpern, Luis Ho, Ed Moran, Wal Sargent, and Joe
Shields over the course of many years. I am grateful to the Committee on
Research at U.C. Berkeley and the conference organizers for providing partial
travel support to attend this meeting (and the latter for their incredible
patience while waiting for my written review!). I also acknowledge the
financial support of NASA grant NAG 5-3556, as well as NASA grants GO-8607 and
GO-8684 from the Space Telescope Science Institute, which is operated by AURA,
Inc., under NASA contract NAS5--26555.

\end{acknowledgments}


\begin{references}

\reference Antonucci, R. R. J. 1993, ARA\&A, 31, 473
\reference Baldwin, J. A., Phillips, M. M., \& Terlevich, R. 1981, 
   PASP, 93, 5
\reference Barth, A. J., Filippenko, A. V., \& Moran, E. C. 1999a, ApJ, 515,
   L61
\reference Barth, A. J., Filippenko, A. V., \& Moran, E. C. 1999b, ApJ, 525,
   673
\reference Barth, A. J., Ho, L. C., Filippenko, A. V., Rix, H.-W., \& Sargent, 
   W. L. W. 2001, ApJ, 546, 205
\reference Barth, A. J., Ho, L. C., Filippenko, A. V., \& Sargent, 
   W. L. W. 1998, ApJ, 496, 133
\reference Barth, A. J., \& Shields, J. C. 2000, PASP, 112, 753
\reference Barth, A. J., et al. 1996, AJ, 112, 1829
\reference Binette, L., Magris, C. G., Stasi\'nska, G., \& Bruzual, A. G. 
   1994, A\&A, 292, 13
\reference Chen, K., Halpern, J. P., \& Filippenko, A. V. 1989, ApJ, 339, 742
\reference Dopita, M. A., \& Sutherland, R. S. 1995, ApJ, 455, 468
\reference Dopita, M. A., \& Sutherland, R. S. 1996, ApJS, 102, 161
\reference Dopita, M. A., et al. 1997, ApJ, 490, 202
\reference Dumont, A. M., \& Collin-Souffrin, S. 1990a, A\&A, 229, 302
\reference Dumont, A. M., \& Collin-Souffrin, S. 1990b, 
   A\&A, 229, 313
\reference Dumont, A. M., \& Collin-Souffrin, S. 1990c, A\&AS, 83, 71
\reference Eracleous, M., \& Halpern, J. P. 1994, ApJS, 90, 1
\reference Fabian, A. C., Arnaud, K. A., Nulsen, P. E. J., \& Mushotzky, R. F.
   1986, ApJ, 305, 9
\reference Falcke, H., Nagar, N. M., Wilson, A. S., \& Ulvestad, J. S.
  2000, ApJ, 542, 197
\reference Ferland, G. J., \& Netzer, H. 1983, ApJ, 264, 105
\reference Filho, M. E., Barthel, P. D., \& Ho, L. C. 2000, ApJS, 129, 93
\reference Filho, M. E., Barthel, P. D., \& Ho, L. C. 2002, A\&A, 385, 425
\reference Filippenko, A. V. 1984, Ph.D. Thesis, California Institute 
   of Technology
\reference Filippenko, A. V. 1985, ApJ, 289, 475
\reference Filippenko, A. V., \& Halpern, J. P. 1984, ApJ, 285, 458
\reference Filippenko, A. V., Ho, L. C., \& Sargent 1993, ApJ, 410, L75
\reference Filippenko, A. V., \& Sargent, W. L. W. 1985, ApJS, 57, 503
\reference Filippenko, A. V., \& Sargent, W. L. W. 1988, ApJ, 324, 134
\reference Filippenko, A. V., \& Sargent, W. L. W. 1992, AJ, 103, 28
\reference Filippenko, A. V., \& Terlevich, R. 1992, ApJ, 397, L79 
\reference Fosbury, R. A. E., Mebold, U., Goss, W. M., \& Dopita, M. A. 1978,
   MNRAS, 183, 549
\reference Gebhardt, K., et al. 2000, ApJ, 539, L73
\reference Halderson, E., Moran, E. C., Filippenko, A. V., \& Ho, L. C.
   2001, AJ, 122, 637
\reference Halpern, J. P., Eracleous, M., Filippenko, A. V., \& Chen, K. 
   1996, ApJ, 464, 704
\reference Halpern, J. P., \& Steiner, J. E. 1983, ApJ, 269, L37
\reference Heckman, T. M. 1980, A\&A, 87, 152
\reference Heckman, T. M. 1981, ApJ, 250, L59
\reference Heckman, T. M., Baum, S. A., van Breugel, W. J. M., \& McCarthy, P.
   1989, ApJ, 338, 48
\reference Ho, L. C. 1999, ApJ, 516, 672
\reference Ho, L. C., Filippenko, A. V., \& Sargent, W. L. W. 1993,
   ApJ, 417, 63
\reference Ho, L. C., Filippenko, A. V., \& Sargent, W. L. W. 1995, 
   ApJS, 98, 477
\reference Ho, L. C., Filippenko, A. V., \& Sargent, W. L. W. 1996,
   ApJ, 462, 183
\reference Ho, L. C., Filippenko, A. V., \& Sargent, W. L. W. 1997a, 
   ApJS, 112, 315
\reference Ho, L. C., Filippenko, A. V., \& Sargent, W. L. W. 1997b, 
   ApJ, 478, 568
\reference Ho, L. C., Filippenko, A. V., \& Sargent, W. L. W. 2003,
   ApJ, in press
\reference Ho, L. C., Filippenko, A. V., Sargent, W. L. W., \& 
   Peng, C. Y. 1997c, ApJS, 112, 391
\reference Ho, L. C., et al. 2000, ApJ, 541, 120
\reference Ho, L. C., et al. 2001, ApJ, 549, L51
\reference Kewley, L. J., Heisler, C. A., Dopita, M. A., \& Lumsden, S. 2001,
   ApJS, 132, 37
\reference Koski, A. T., \& Osterbrock, D. E. 1976, ApJ, 203, L49
\reference Krolik, J. H. 1999, Active Galactic Nuclei (Princeton:
   Princeton Univ. Press)
\reference Maoz, D., et al. 1995, ApJ, 440, 91
\reference Maoz, D., et al. 1998, AJ, 116, 55
\reference Nagar, N. M., Falcke, H., Wilson, A. S., \& Ho, L. C. 2000,
   ApJ, 542, 186
\reference Rix, H.-W., \& White, S. D. M. 1992, MNRAS, 254, 389
\reference Roberts, T. P., \& Warwick, R. 2000, MNRAS, 315, 98
\reference Sabra, B. M., Shields, J. C., \& Filippenko, A. V. 2000,
   ApJ, 545, 157
\reference Sabra, B. M., Shields, J. C., Ho, L. C., Barth, A. J., \& 
   Filippenko, A. V. 2003, ApJ, in press
\reference Shields, J. C. 1992, ApJ, 399, L27
\reference Shields, J. C., \& Filippenko, A. V. 1990, ApJ, 353, L7
\reference Shields, J. C., et al. 2000, ApJ, 534, L27
\reference Taniguchi, Y., Shioya, Y., \& Murayama, T. 2000, AJ, 120, 1265
\reference Terashima, Y., Ho, L. C., \& Ptak, A. F. 2000, ApJ, 539, 161
\reference Terlevich, R., \& Melnick, J. 1985, MNRAS, 213, 841
\reference Veilleux, S., \& Osterbrock, D. E. 1987, ApJ, 63, 295

\end{references}
\end{document}